\documentclass[pra,aps,twocolumn,superscriptaddress,showpacs,amsmath,amssymb,floatfix]{revtex4}

\usepackage{graphics}

\begin{document}

\title{Critical velocity of superfluid flow through \\
single barrier and periodic potentials }

\author{Gentaro Watanabe}
\affiliation{CNR INFM-BEC and Department of Physics, University of Trento, 38050 Povo, Italy}
\affiliation{RIKEN, 2-1 Hirosawa, Wako, Saitama 351-0198, Japan}
\author{F. Dalfovo}
\affiliation{CNR INFM-BEC and Department of Physics, University of Trento, 38050 Povo, Italy}
\author{F. Piazza}
\affiliation{CNR INFM-BEC and Department of Physics, University of Trento, 38050 Povo, Italy}
\author{L.~P. Pitaevskii}
\affiliation{CNR INFM-BEC and Department of Physics, University of Trento, 38050 Povo, Italy}
\affiliation{Kapitza Institute for Physical Problems, 119334 Moscow, Russia}
\author{S. Stringari}
\affiliation{CNR INFM-BEC and Department of Physics, University of Trento, 38050 Povo, Italy}

\date{July 2, 2009}

\begin{abstract}
We investigate the problem of an ultracold atomic gas in the
superfluid phase flowing in the presence of a potential barrier or
a periodic potential. We use a hydrodynamic scheme in the local
density approximation (LDA) to obtain an analytic expression for
the critical current as a function of the barrier height or the
lattice intensity, which applies to both Bose and Fermi
superfluids. In this scheme, the stationary flow becomes
energetically unstable when the local superfluid velocity is equal
to the local sound velocity at the point where the external
potential is maximum. We compare this prediction with the results
of the numerical solutions of the Gross-Pitaevskii and
Bogoliubov-de Gennes equations. We discuss the role of long
wavelength excitations in determining the critical velocity. Our
results allow one to identify the different regimes of superfluid
flow, namely, the LDA hydrodynamic regime, the regime of
quantum effects beyond LDA for weak barriers and the regime of
tunneling between weakly coupled superfluids for strong barriers.
We finally discuss the relevance of these results in the context
of current experiments with ultracold gases.
\end{abstract}

\maketitle

\section{Introduction}

The critical velocity of superfluid flow is one of the most fundamental
issues in the physics of quantum fluids. A mechanism for the onset
of dissipation is provided by Landau instability \cite{landau,
nozieres_pines,pethick_smith,book}: if the excitation spectrum
satisfies suitable criteria, there exists a critical flow velocity
above which the kinetic energy of the superfluid can be dissipated
{\it via} the creation of excitations. For uniform weakly
interacting Bose gases, this critical velocity $v_c$ coincides
with the velocity of sound waves. Several experiments have already
been devoted to observe and characterize the onset of dissipation
in superfluids made of ultracold atomic gases, including studies
of Bose-Einstein condensates (BECs) in relative motion with
respect to external perturbations, such as point-like defects
\cite{mit_old} or weak optical lattices \cite{lens2}.

The present work was stimulated by the recent experiment by Miller
{\it et al.} \cite{Miller}, where a one-dimensional (1D) optical
lattice is used to measure the critical velocity of the superfluid
flow of ultracold fermionic atoms in the BCS-BEC crossover. The
lattice is produced by two intersecting laser beams focused in the 
central part of the trapped gas. A frequency difference between the
two beams is used to produce a relative motion of the lattice and
the gas. The number of fermion pairs which remain in the superfluid 
after the application of the moving lattice is measured as a function 
of the relative velocity in order to determine the critical velocity
for the onset of dissipation.  The experiment confirms a theoretical 
prediction that superfluidity is most robust at unitarity 
\cite{andrenacci,sensarma,combescot}. It also shows that the critical 
velocity is very sensitive to the intensity of the lattice. This 
shares some similarity with the flow of a
superfluid in the presence of a single potential barrier. In the
case of ultracold fermions in the BCS-BEC crossover, the latter
problem has been recently addressed in the framework of the
Bogoliubov-de Gennes (BdG) theory by Spuntarelli {\it et al.}
\cite{camerino}, who also found a strong dependence of the
critical velocity on the barrier height. These works (see also,
e.g., Refs.~\cite{orso,burkov,ancilotto,yunomae,ganesh}) are recent
fermionic examples of a wider field of investigations of critical
current phenomena mainly studied, in the past, for Bose-Einstein
condensed gases. We notice however that the similarities and
differences between the behavior of the critical flow in the
presence of a single barrier and of a periodic potential, as well
as the role of quantum statistics have never been addressed in a
clear and systematic way.

Our work is aimed to establish an appropriate framework where 
one can compare the different situations (bosons vs. fermions and 
single barrier vs. optical lattice) and extract useful indications 
for available and/or feasible experiments. To this purpose we first 
introduce the hydrodynamic formalism in the local density approximation 
(LDA) and we assume a polytropic equation of state, which applies both to 
a Fermi superfluid at unitarity (i.e., when the $s$-wave scattering 
length $a_s$ is much larger than the interparticle distance, which 
in turn is much larger than the range of the interatomic potential)
and to a Bose-Einstein condensate. In this way we derive a universal 
relation between the critical current for energetic instability and 
the maximum value of the external potential. We then compare the LDA 
hydrodynamic prediction with the results of the numerical solutions 
of the Gross-Pitaevskii (GP) and Bogoliubov-de Gennes equations for 
bosons and fermions, respectively. This comparison allows us to 
identify different regimes of superfluid flow depending on the 
relevant energy and length scales of the system: a regime of 
hydrodynamic flow in LDA, a regime of macroscopic flow beyond LDA,
and a regime of weakly coupled superfluids separated by thin and 
strong barriers. From our analysis one can see that the BdG results 
of Ref.~\cite{camerino} for fermions at unitarity nicely fall into 
the LDA hydrodynamic regime. Conversely, the parameters of the 
experiment in Ref.~\cite{Miller} are such that the LDA is not applicable,
the healing length being of the same order of the lattice spacing.
Indeed we find that the experimental results are not reproduced 
by the LDA, as expected, but we find that they also disagree with the BdG 
theory in a periodic potential with parameters similar to the ones
of the experiment. This disagreement suggests that the interpretation 
of the experimental observations in terms of energetic instability of
stationary superfluid flow in a uniform 1D periodic potential remains
an open issue.

\section{Hydrodynamics in the local density approximation}
\label{sect-hydro}

 Our starting point is a hydrodynamic theory in the local density
approximation at zero temperature. Let us consider a superfluid
which extends to infinity in three dimensions and is subject to a
1D potential, $V_{\rm ext} (z)$, which has a
finite maximum value $V_{\rm max}$. Let the superfluid be in a
stationary configuration characterized by the density profile
$n(z)$ and the (quasi-) momentum $P(z)$ along the $z$-direction.
The LDA assumes that, locally, the system behaves like a uniform
gas; thus the energy density $e(n,P)$ can be written in the form
$e(n,P)=nP^2/2m + e(n,0)$ and one can define the local chemical
potential $\mu(n)=\partial e(n,0) /\partial n$. The density
profile of the gas at rest in the presence of the external
potential can be obtained from the Thomas-Fermi relation $\mu_0 =
\mu(n(z)) + V_{\rm ext} (z)$. If the gas is flowing with a
constant current density $j=n(z)v(z)$, the Bernoulli equation for
the stationary velocity field $v(z)$ is
\begin{equation}
\mu_j = \frac{m}{2} \left[ \frac{j}{n(z)}\right]^2 + \mu(n) +
V_{\rm ext} (z),
\label{eq:mu}
\end{equation}
where $\mu_j$ is the $z$-independent value of the chemical
potential. This equation gives the density profile, for any given
current $j$, once the equation of state $\mu(n)$ of the uniform
gas of density $n$ is known. We assume the uniform gas to obey a
polytropic equation of state, $\mu(n)=\alpha n^\gamma$, and we
consider two cases: i) a dilute Bose gas with repulsive interaction,
for which one has $\gamma=1$ and $\alpha=g=4\pi\hbar^2a_s/m$, where
$g$ is the interaction parameter, $m$ is the atom mass, and $a_s>0$
is the $s$-wave scattering length; ii) a dilute Fermi gas at
unitarity, for which $\mu(n)=(1+\beta) E_{\rm F}$,
where $\beta$ is a universal parameter \cite{giorgini,beta} while
$E_{\rm F} = \hbar^2k_{\rm F}^2/(2m)$ and $k_{\rm F}=(3\pi^2 n)^{1/3}$
are the Fermi energy and momentum, respectively, of a uniform
non-interacting Fermi gas of density $n$, so that
$\gamma=2/3$ and $\alpha=(1+\beta) (3\pi^2)^{2/3}\hbar^2/(2m)$.
Using the equation of state one can write 
\begin{equation}
m c_s^2(z) = n \frac{\partial}{\partial n} \mu(n) = \gamma \mu(n) \, ,
\label{eq:mc2}
\end{equation}
where $c_s(z)$ is the local sound velocity, which depends on $z$ 
through the density profile $n(z)$. In a uniform gas of density $n_0$, 
the sound velocity is given by $c_s^{(0)}= [\gamma\mu(n_0)/m]^{1/2}$.

In the framework of LDA, the system becomes energetically
unstable when the local superfluid velocity, $v(z)$ at some point
$z$ is equal to the local sound velocity, $c_s(z)$. If the 
external potential has a maximum at $z=z_0$ [i.e., $V_{\rm ext}(z_0)
= V_{\rm max}$], then at the same point the density is minimum, 
$c_s(z)$ is minimum and $v(z)$ is maximum. This means that the 
superfluid becomes first unstable precisely at $z=z_0$. The 
condition for the occurrence of the instability can be written as 
$m[j_c/n_c(z_0)]^2=\gamma \mu[n_c(z_0)]= \gamma \alpha n_c^\gamma(z_0)$, 
where $n_c(z)$ is the density profile calculated at the critical 
current \cite{kink}. By inserting this condition into the Bernoulli 
equation (\ref{eq:mu}),
after a straightforward calculation one obtains the following
implicit relation for the critical current:
\begin{equation}
j_c^2 = \frac{\gamma}{m \alpha^{2/\gamma}}
\left[ \frac{2\mu_{j_c}}{2+\gamma}
\left(1-\frac{V_{\rm max}}{\mu_{j_c}}\right)
\right]^{\frac{2}{\gamma}+1} \, .
\label{eq:jc}
\end{equation}
It is worth noticing that this equation contains only $z$-independent
quantities. It is also independent of the shape of the external
potential: the only relevant parameter being its maximum value
$V_{\rm max}$. Moreover, it can be applied to both bosons and
fermions. Its version for bosons in slowly varying potentials
was already discussed in Refs.~\cite{mamaladze,hakim}
(see also Ref.~\cite{leboeuf}).

\begin{figure}[htbp]
\begin{center}\vspace{0.0cm}
  \rotatebox{0}{ \resizebox{8.2cm}{!}
     {\includegraphics{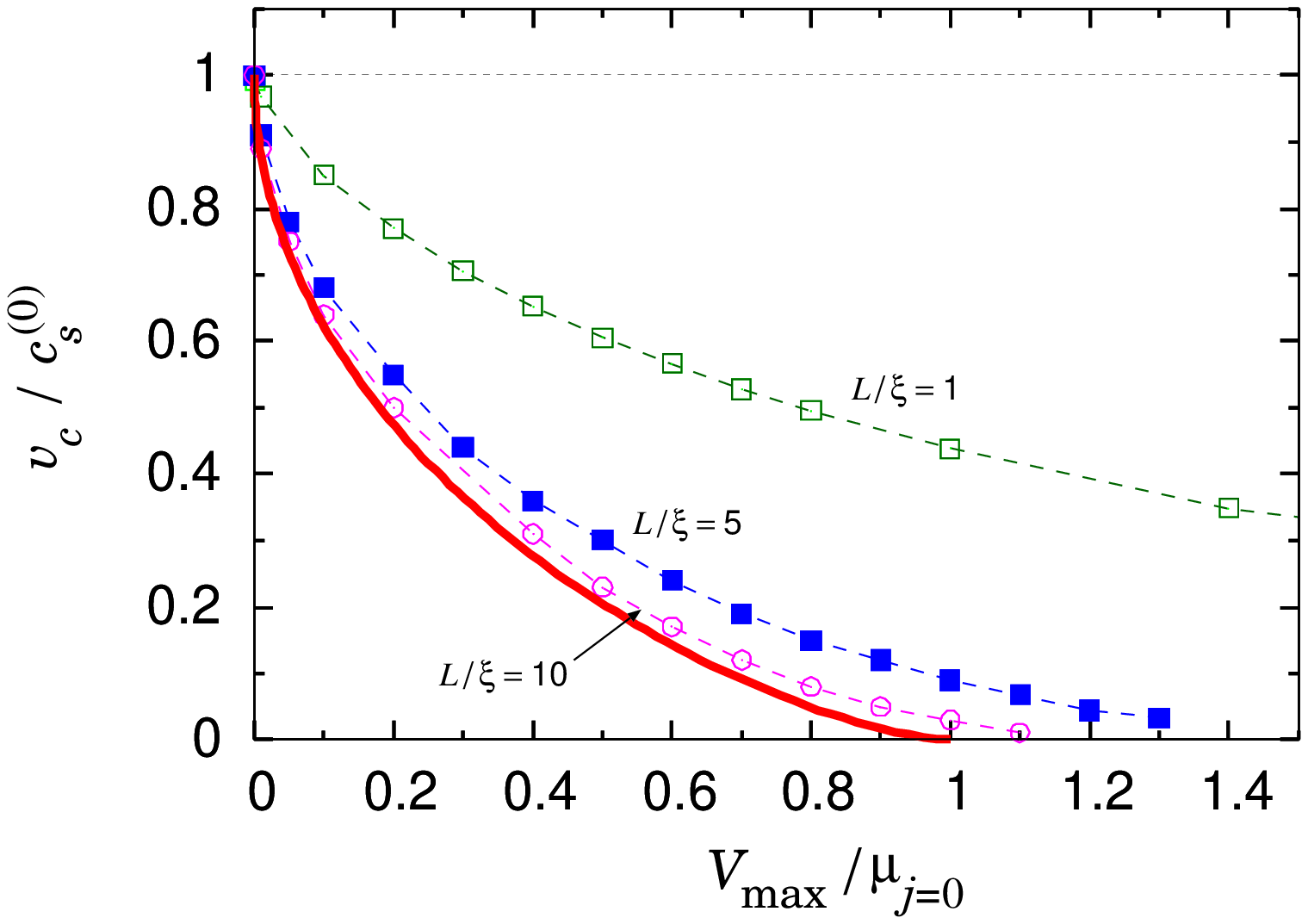}}}
   \rotatebox{0}{\resizebox{8.2cm}{!}
    {\includegraphics{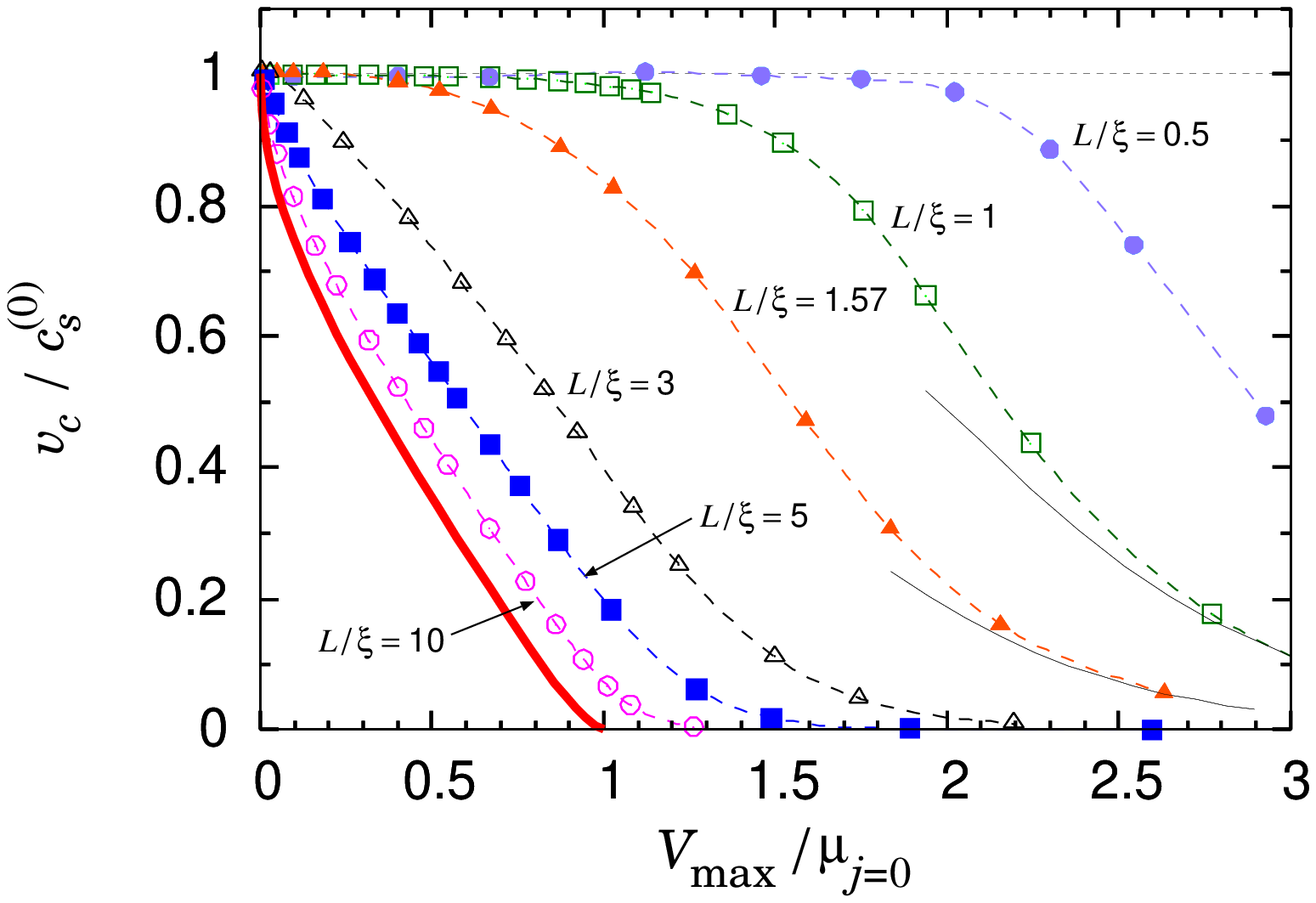}}}
 \caption{\label{fig-bose}(Color online)\quad Critical velocity
for energetic instability of a dilute bosonic superfluid subject to
a 1D external potential. The critical velocity is given in units
of the sound velocity of a uniform gas, $c_s^{(0)}$,  and is plotted
as a function of the maximum of the external potential in units of the
chemical potential $\mu_{j=0}$ of the superfluid at rest. Top panel:
the case of a single potential barrier. Bottom panel: the case of a
periodic potential. Thick solid lines: prediction of the hydrodynamic
theory within the LDA, calculated from Eq.~(\ref{eq:jc}). Symbols:
results obtained from the numerical solution of the GP equation
(\ref{eq:gp}) for various values of $L/\xi$.
Filled circles: $L/\xi=0.5$ ($gn_0/E_{\rm R}=0.1$);
open squares: $L/\xi=1$ ($gn_0/E_{\rm R}=0.4$);
filled triangles: $L/\xi=1.57$ ($gn_0/E_{\rm R}=1$);
open triangles: $L/\xi=3$ ($gn_0/E_{\rm R}=3.65$);
filled squares: $L/\xi=5$ ($gn_0/E_{\rm R}=10$);
open circles: $L/\xi=10$ ($gn_0/E_{\rm R}=40$).
The thinner black solid lines are the tight-binding prediction
(\protect\ref{eq:tb}) for $L/\xi=1$ and $1.57$.
Dashed lines are guides to the eye.
}
\end{center}
\end{figure}

\begin{figure}[htbp]
\begin{center}\vspace{0.0cm}
  \rotatebox{0}{\resizebox{8.2cm}{!}
    {\includegraphics{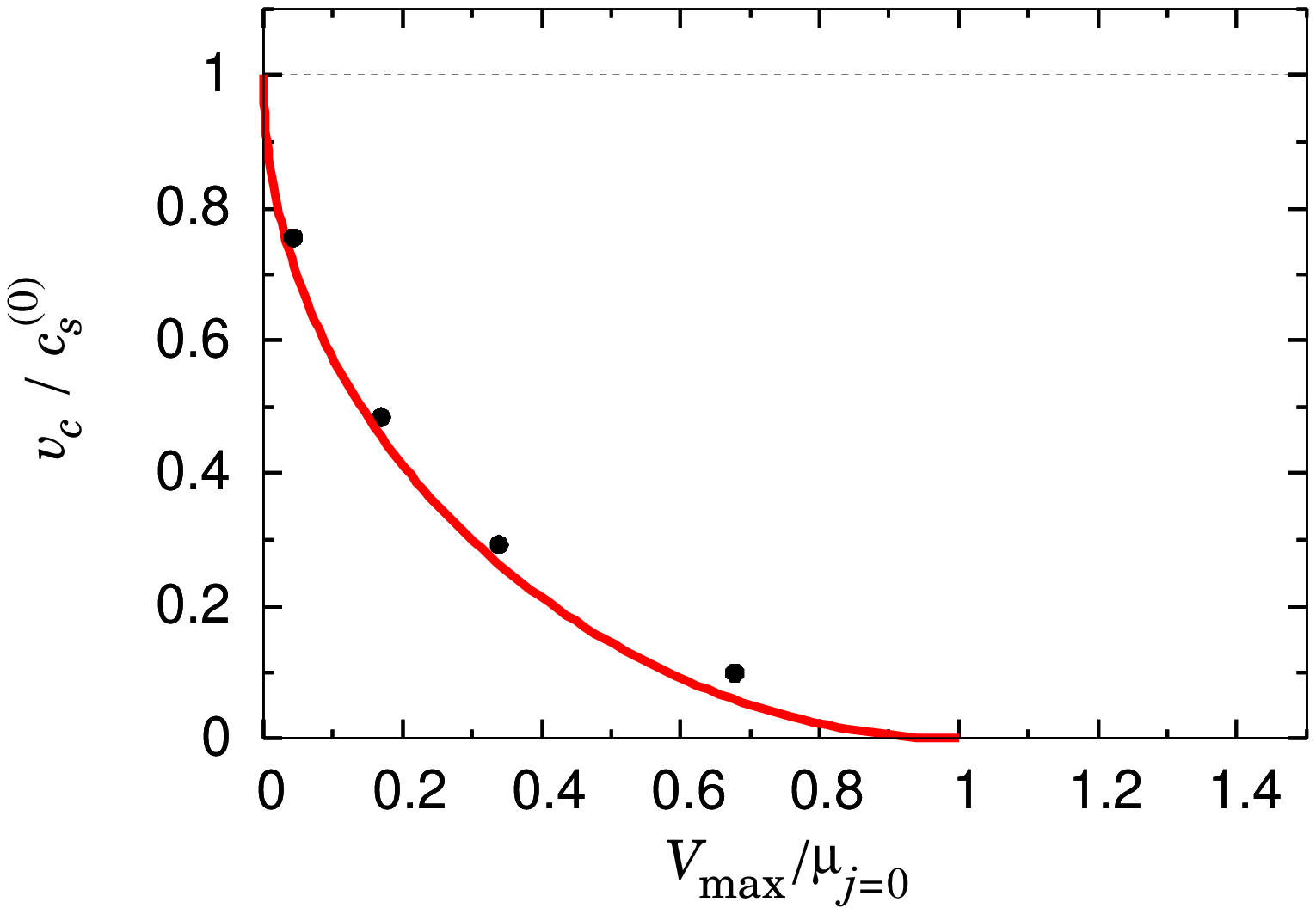}}}
  \rotatebox{0}{\resizebox{8.2cm}{!}
    {\includegraphics{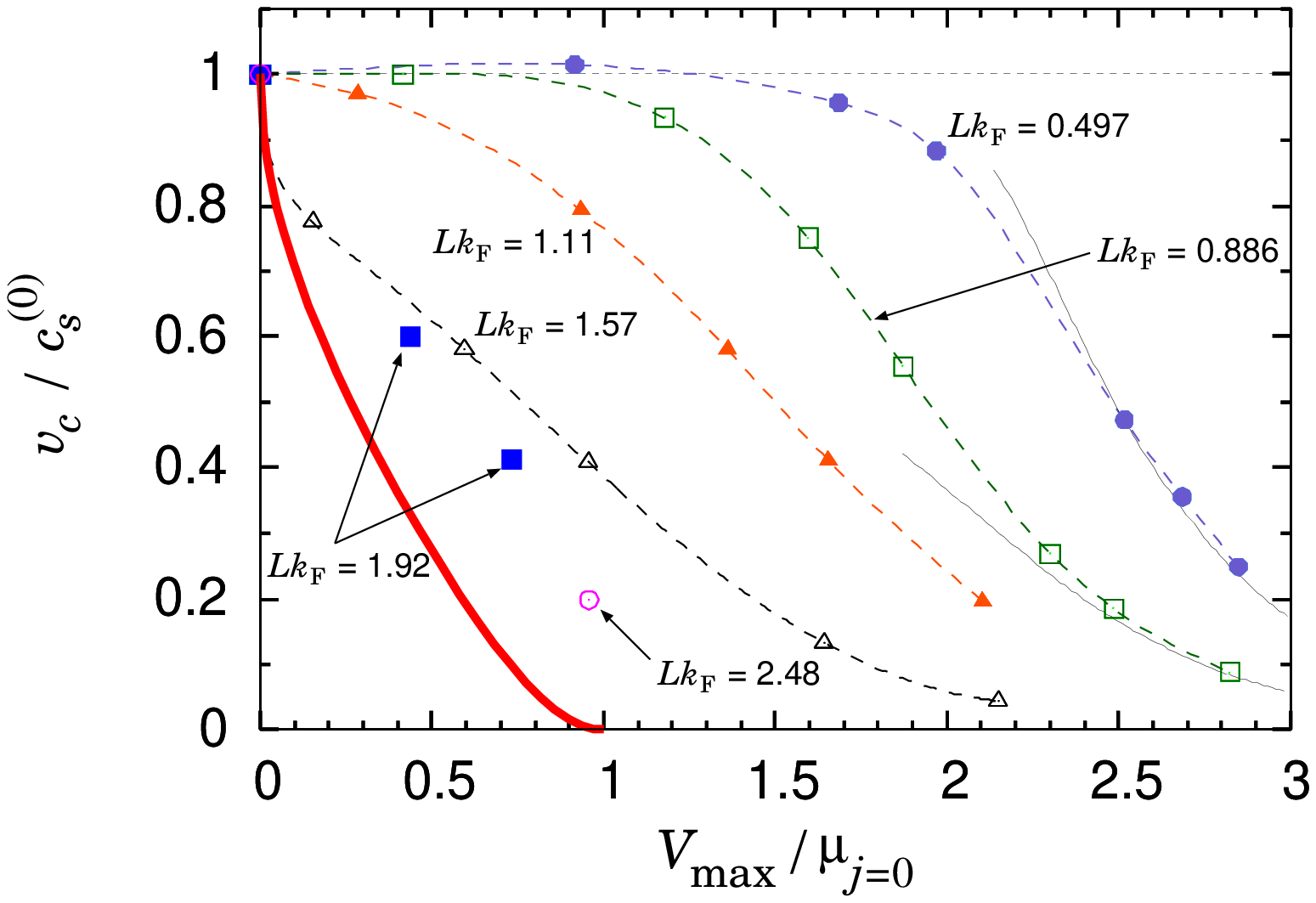}}}
\caption{\label{fig-fermi}(Color online)\quad Same as in
Fig.~\protect\ref{fig-bose} but for a superfluid of dilute fermions
at unitarity. Thick solid lines: prediction of the LDA hydrodynamic 
theory, from Eq.~(\ref{eq:jc}). Symbols in the top panel: BdG results
of Ref.~\protect\cite{camerino} with $Lk_{\rm F}=4$. Symbols in the
bottom panel represent our BdG results for a periodic lattice. The 
$s\neq 0$ values are shown as follows:
filled circles for $Lk_{\rm F}=0.497$ ($E_{\rm F}/E_{\rm R}=0.1$) and
$s=0.1$, $0.5$, $1$, $2.5$, $3$, and $3.5$;
open squares for  $Lk_{\rm F}=0.886$ ($E_{\rm F}/E_{\rm R}=0.3178$) and
$s=0.1$, $0.5$, $1$, $1.5$, $2.5$, $3$, and $4$;
filled triangles for $Lk_{\rm F}=1.11$ ($E_{\rm F}/E_{\rm R}=0.5$) and
$s=0.1$, $0.5$, $1$, $1.5$, and $2.5$;
open triangles for $Lk_{\rm F}=1.57$ ($E_{\rm F}/E_{\rm R}=1$) and
$s=0.1$, $0.5$, $1$, $2.5$, and $4$;
filled squares for $Lk_{\rm F}=1.92$ ($E_{\rm F}/E_{\rm R}=1.5$ and
$s=0.5$ and $1$;
open circle for  $Lk_{\rm F}=2.48$ ($E_{\rm F}/E_{\rm R}=2.5$) and
$s=2.5$.
The result for $Lk_{\rm F}=0.497$ and $s=0.1$ being larger than unity 
is due to the numerical uncertainty, which is estimated to be 
of the order of $3$\% \ or less for all BdG points in the bottom panel 
of this figure. The thinner black solid lines are the tight-binding 
prediction (\protect\ref{eq:tb}) for $Lk_{\rm F}=0.497$ and $0.886$. 
Dashed lines are guides to the eye.
}
\end{center}
\end{figure}

In Figs.~\ref{fig-bose} and \ref{fig-fermi}, we plot the critical
velocity obtained from the hydrodynamic expression (\ref{eq:jc})
for bosons and fermions, respectively (thick solid lines). The
upper plots refer to the case of a single 1D rectangular barrier
of width $L$ and height $V_{\rm max}$, the superfluid having
an asymptotic constant density $n_0$ 
and velocity $v_0\equiv j/n_0$
at large distances from the
barrier. The bottom plots refer to the case of a 1D periodic
potential $V_{\rm ext}(z)=V_{\rm max} \sin^2{(q_{\rm B} z)} =
sE_{\rm R}\sin^2{(q_{\rm B} z)}$, where $q_{\rm B}$ is the Bragg
wave vector, $E_{\rm R}\equiv\hbar^2 q_{\rm B}^2/2m$ is the recoil
energy, $s$ is the lattice strength, and the superfluid has an
average density $n_0$ and a macroscopic velocity defined by 
$v_0\equiv j/n_0$. 
In all cases, the critical velocity
$v_c=j_c/n_0$ is normalized to the value of the sound velocity in
the uniform gas, $c_s^{(0)}$, and is plotted as a function of
$V_{\rm max}/\mu_{j=0}$ (see Appendix \ref{app-lda}). 
The limit $V_{\rm max}/\mu_{j=0} \to 0$ corresponds to the usual Landau 
criterion for a uniform superfluid flow in the presence of a small 
external perturbation, i.e., a critical velocity equal to the sound
velocity of the gas. In this hydrodynamic scheme, as mentioned before,
the critical velocity decreases when $V_{\rm max}$ increases
mainly because the density has a local depletion and the velocity
has a corresponding local maximum, so that the Landau instability
occurs earlier. When $V_{\rm max} = \mu_{j_c}$, the density
exactly vanishes below the barrier and hence the critical velocity
goes to zero.

The condition for the applicability of the LDA expression (\ref{eq:jc})
is that the typical length scale of the external potential must be 
much larger than the healing length $\xi$ of the superfluid. For a 1D 
potential barrier of width $L$, this implies $L/\xi \gg 1$ \cite{smoothing}. 
In an optical 
lattice, the typical scale is the lattice spacing $d=\pi/q_{\rm B}$; in 
order to compare the two cases we define the barrier width $L$ as half 
the lattice spacing, i.e., $L\equiv d/2=\pi/(2q_{\rm B})$. The healing 
length of a dilute Bose superfluid of density $n_0$ is $\xi=\hbar/
(2mgn_0)^{1/2}$, while for a Fermi gas at unitarity one has 
$\xi \sim 1/k_{\rm F}$, which is consistent
with the BCS coherence length $\xi_{\rm BCS}=\hbar v_{\rm F}/
\Delta_{\rm gap}$, where $v_{\rm F}=\hbar k_{\rm F}/m$ and
$\Delta_{\rm gap}$ is the pairing gap, which is of order 
$E_{\rm F}$ at unitarity. Effects beyond LDA become
important when $\xi$ is of the same order or larger than $L$; they
cause a smoothing of both density and velocity distributions, as well
as the emergence of solitonic excitations, which are expected to
play an important role in determining the critical velocity. This is
well known for bosons, where a supercritical current results in the
emission of shock waves in classical hydrodynamics and of solitons
in the Gross-Pitaevskii theory \cite{hakim,pavloff,lesz}.

\begin{figure}[htbp]
\begin{center}\vspace{0.0cm}
  \rotatebox{0}{\resizebox{7.5cm}{!}
    {\includegraphics{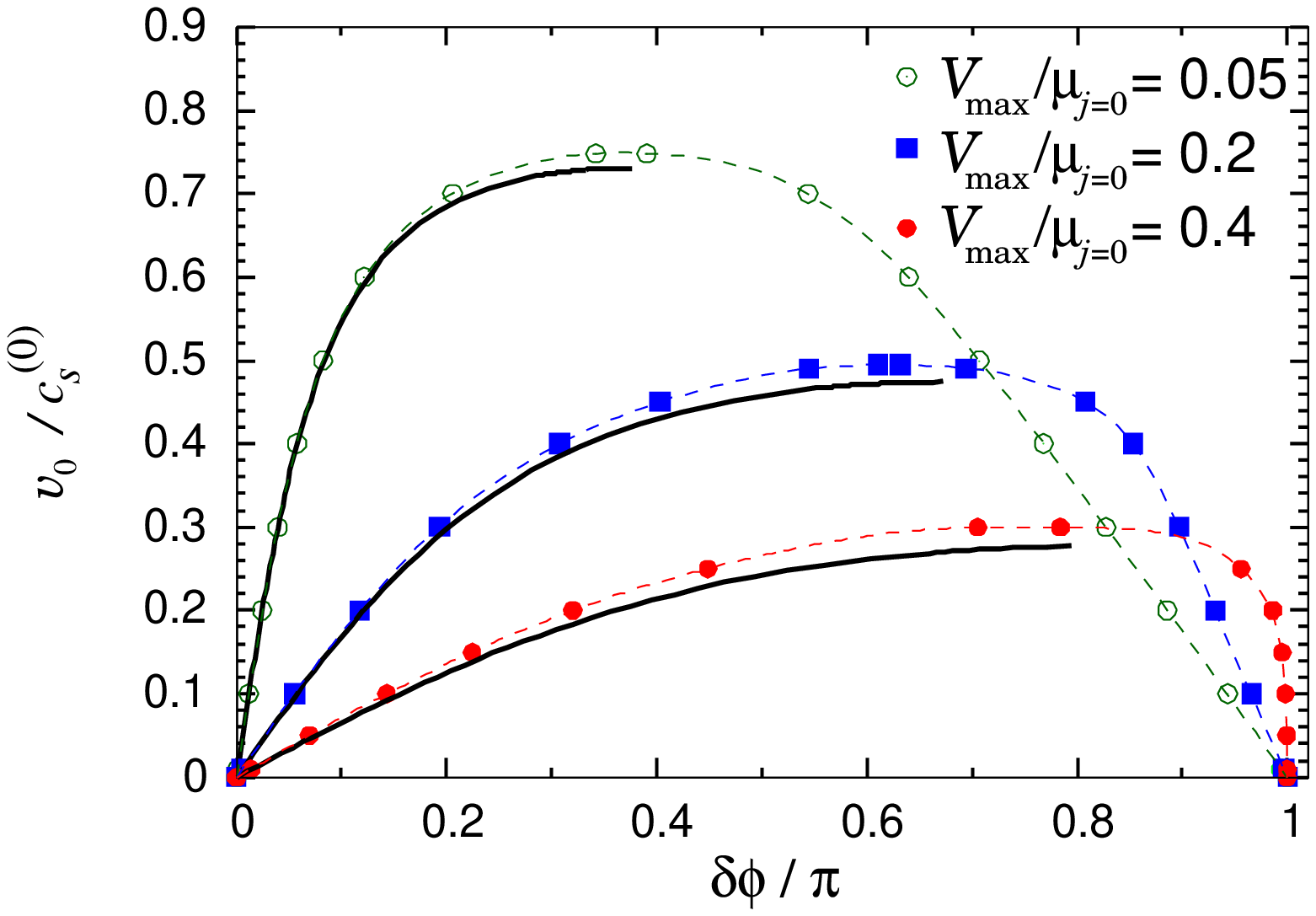}}}
  \rotatebox{0}{\resizebox{7.5cm}{!}
    {\includegraphics{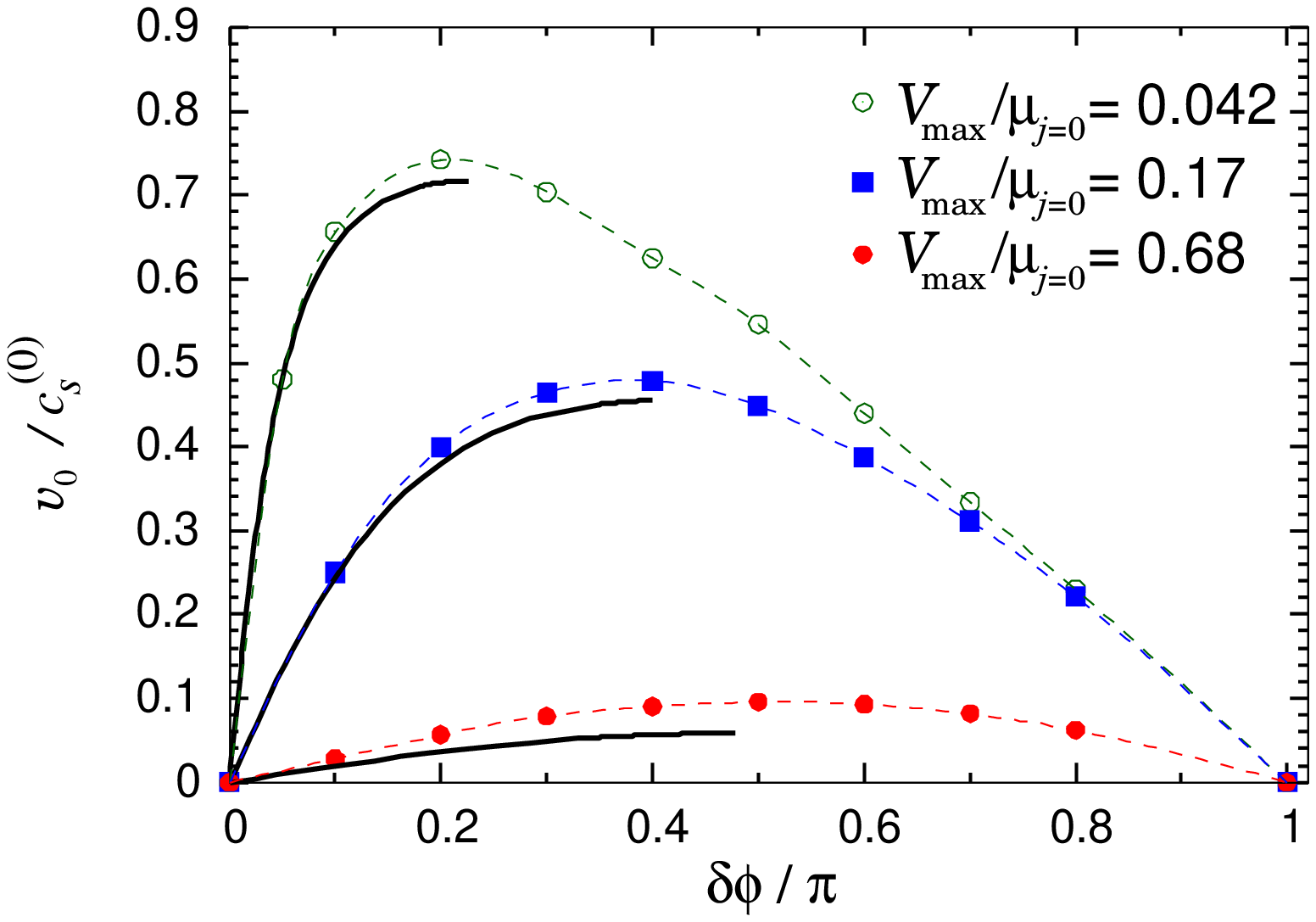}}}
\caption{\label{fig-iphi}(Color online)\quad
Velocity-phase diagrams in the LDA hydrodynamic regime
for a bosonic superfluid (top panel) and for a fermionic
superfluid at unitarity (bottom panel) through a single 
barrier. The thick solid lines show the prediction of the 
LDA hydrodynamic theory.  The symbols in the top panel are 
the results obtained from the GP equation; those in the bottom 
panel are taken from the BdG results of Ref.~\cite{camerino}. 
The barrier heights for bosonic case are 
$V_{\rm max}/\mu_{j=0}=0.05$ (open circles), $0.2$ (filled squares), 
and $0.4$ (filled circles). Those for fermionic case are $V_{\rm
max}/\mu_{j=0}=0.042$ (open circles), $0.17$ (filled squares), and
$0.68$ (filled circles). The barrier width is $L/\xi=10$ for the 
bosonic case, and $Lk_{\rm F}=4$ for the fermionic case; these 
parameters are such that the corresponding curves for $v_c/c_s^{(0)}$ 
in the top panels of Figs.~\ref{fig-bose} and \ref{fig-fermi} are
qualitatively similar. The dashed lines are guides to the eye.
}
\end{center}
\end{figure}

In closing this section, it is worth mentioning that within the
LDA scheme one can easily calculate also the
current(velocity)-phase relation of the superfluid flowing through
a single potential barrier. The phase $\phi$ of the order
parameter is related to the local velocity by $v(z)=(\hbar/M)\nabla
\phi$, where $M=m$ for Bose superfluids and $M=2m$ for Fermi
superfluids. For a given asymptotic velocity $v_0=j/n_0$ at $z = \pm
\infty$, one can directly integrate the phase across the barrier
in order to obtain the phase difference $\delta \phi$. By 
subtracting the contribution of $v_0$, the phase difference can be 
defined as $\delta\phi\equiv (M/\hbar)\int^{\infty}_{-\infty}dz\, 
[v(z)-v_0]$. In the LDA, the local velocity $v(z)$ is 
constant under the rectangular barrier and one thus obtains 
$\delta\phi  = ML[v(z_0)-v_0]/\hbar$. Examples are shown
in Fig.~\ref{fig-iphi} for bosons with $L/\xi=10$ (top panel) and
unitary fermions with $Lk_F=4$ (bottom panel). The predictions of
the hydrodynamic theory within LDA are shown as solid lines for
three values of the barrier height. Each line is drawn up to the
value of $\delta \phi_c$ for which the velocity becomes maximum.
These lines are called stable branches in the
current(velocity)-phase diagram; on the right of the maximum there
are no solutions within LDA (see the end of Section \ref{sect-gp}
for details). The maximum velocity coincides with the critical
velocity $v_c$ plotted in the top panels of Figs.~\ref{fig-bose}
and \ref{fig-fermi}. 

\section{Gross-Pitaevskii theory for bosons}
\label{sect-gp}

In order to account for effects beyond the LDA hydrodynamic theory
we use the GP theory for dilute bosons \cite{gross,pitaevskii,gp}
and the BdG equations for fermions at unitarity \cite{bdg}. Let us 
first consider the case of a dilute 3D Bose-Einstein condensate at 
zero temperature subject to either a 1D rectangular potential 
barrier or a periodic potential. The system is well described by 
the 1D GP equation \cite{gp}:
\begin{equation}
- \frac{\hbar^2}{2m}\partial_{z}^2 \Psi + V_{\rm ext}(z) \Psi +
g | \Psi |^2 \Psi = \mu \Psi ,
\label{eq:gp}
\end{equation}
where $\Psi(z)=\sqrt{n(z)}\exp[i\phi(z)]$ is the complex order
parameter of the condensate, $\phi(z)$ is its phase, and $\mu$ is
the chemical potential. The local superfluid velocity is given by
$v(z)=(\hbar/m)\partial_z \phi(z)$ and the sound velocity in a 
uniform condensate of density $n_0$ is $c_s^{(0)}=\sqrt{gn_0/m}$.
We numerically solve the GP equation in order to find the condition
for the energetic stability of the condensate.

In the case of a single barrier, we calculate the critical current 
by looking for solutions of the time-independent GP equation 
(\ref{eq:gp}) at a fixed current $j$ and for a given asymptotic 
density $n_0$ \cite{solvingGPE}. We also use the time-dependent 
version of the GP equation in order to check the dynamical behavior 
of the unstable solutions. In the case of the periodic potential, 
the critical current is instead found by using the Bloch wave
formalism and solving the linear stability problem for the GP energy
functional as in Refs.~\cite{wuandniu,machholm,modugno}. For the same
case, we also use an alternative method, namely, a hydrodynamic analysis
of the excitations, which is valid when the excitations that trigger
the instability are those with wavelength much larger than the lattice
spacing. For a superfluid with average density $n_0$ and quasi-momentum
$P$, this approach predicts the dispersion relation
\begin{equation}
\omega (q) =\frac{\partial^2 e}{\partial n_0 \partial P}\ q
+\sqrt{\frac{\partial^2 e}{\partial n_0^2}
\frac{\partial^2 e}{\partial P^2}}\ |q|\ ,
\label{eq:dispersion}
\end{equation}
where $\hbar\omega$ and $q$ are the energy and the wavenumber
of the excitations. The energetic instability occurs when
\cite{machholm,taylor,orso}
\begin{equation}
\frac{\partial^2 e}{\partial n_0 \partial P} =
\sqrt{\frac{\partial^2 e}{\partial n_0^2}
\frac{\partial^2 e}{\partial P^2}}\ .
\label{eq:vcrit-hydro}
\end{equation}
In practice, we calculate the energy density $e(n_0,P)$ by solving
the GP equation for a given average density $n_0$ and quasi-momentum
$P$ of the superfluid. The value of $P$ for which Eq.~(\ref{eq:vcrit-hydro})
is satisfied is the critical quasi-momentum, $P_c$. Finally, the critical
velocity can be calculated by using the relation $v_c= (1/n_0)
(\partial e / \partial P )_{P_c}$ (see Appendix B for details).
We have checked that the critical velocity obtained in this way agrees
with that obtained by means of the complete linear stability analysis
within 1\% in the whole range of $gn_0/E_{\rm R}$ and $V_{\rm max}$
considered in the present work. This confirms that the energetic
instability in the periodic potential is driven by long-wavelength
excitations, since the excitation energy of the sound mode is the smallest 
in this limit, as it occurs in the case of a single barrier. Equation
(\ref{eq:vcrit-hydro}) can be used also for unitary fermions, as we 
will see in Section IV.   

In Fig.~\ref{fig-bose}, we show the critical velocity in the case
of the single rectangular barrier (top panel) and the periodic
potential (bottom panel) for various values of $L/\xi$. In both
cases, one clearly sees that the results of the GP equation
approach the LDA prediction for $L/\xi\gg 1$, as expected. The 
way of approach is, however, different. 
In the case of the periodic potential, $v_c$
exhibits a plateau for $L/\xi \alt 1$ and small $V_{\rm max}$; the
plateau is instead absent in the case of the single barrier.
The latter case is simpler, because the chemical potential and the
excitation spectrum are fixed by the asymptotic density $n_0$ only
and are unaffected by the external potential. In the limit of weak 
and thin barriers, our GP results agree with the analytic expression 
$v_c/c_s^{(0)} \simeq 1-(3/4) (LV_{\rm max}/\xi gn_0)^{2/3}$
already derived in Ref.~\cite{hakim}; conversely, for thick barriers 
($L/\xi \gg 1$), the $V_{\max} \to 0$ limit of the LDA expression 
(\ref{eq:jc}) is $v_c/c_s^{(0)}\simeq 1-\sqrt{3/2}\, (V_{\rm
max}/gn_0)^{1/2}$ \cite{lesz}. All curves obtained from the GP 
equation for the single barrier have a curvature which lies in 
between these two limiting cases. Instead, the periodic potential
extends over the whole system and, consequently, for a given average 
density $n_0$, a change of $V_{\rm max}$ affects both $\mu$  and 
the excitation spectrum in a way that depends also on the value of 
the healing length $\xi$. In particular, if $\xi$ is larger than the 
lattice spacing and $V_{\rm max}$ is not too large, the energy 
associated with quantum pressure, which is proportional to $1/\xi^2$,  
acts against local deformations of the order parameter and the latter 
remains almost unaffected by the modulation of the external 
potential. This is the regime of the plateau in Fig.~\ref{fig-bose}. 
In terms of Eq.~(\ref{eq:vcrit-hydro}), this regime occurs when the 
left-hand-side is $\simeq P/m$ and the right-hand-side 
is $\simeq c_s^{(0)}$, so that the critical quasi-momentum obeys 
the relation $P_c/m=c_s^{(0)}$, which is the usual Landau criterion 
for a uniform superfluid in the presence of weak impurities 
(see Appendix \ref{app-pc} for more details).  This regimes ends when, 
by increasing $V_{\rm max}$, the sound velocity $c_s$ in the lattice 
[i.e., the right-hand-side of Eq.~(\ref{eq:vcrit-hydro}) evaluated 
at $P=0$] becomes comparable or less than the maximum of the 
left-hand-side. Since, for small $V_{\rm max}$, $\mu$ is proportional 
to $n_0$ and the maximum of the left-hand-side of Eq.~(\ref{eq:vcrit-hydro}) 
is of the order of $q_{\rm B}/m$, the above condition is obtained for  
$\mu \simeq E_{\rm R}$. When $V_{\rm max}$ is further increased, the
chemical potential $\mu$ becomes larger than $E_{\rm R}$, the density
is forced to oscillate and $v_c/c_s^{(0)}$ starts decreasing . The 
system eventually reaches a regime of weakly coupled superfluids 
separated by strong barriers. This regime will be discussed in 
Section \ref{tb}.

It is worth noticing that, for $V_{\rm max} < 2gn_0$ \cite{diakonov},
loops (``swallow tails") appear in the Bloch band structure of the 
superfluid in the periodic potential as a consequence of the nonlinearity 
of the GP equation. These states and their stability have been deeply 
discussed in Ref.~\cite{machholm} and are accounted for in our 
calculations. The sets of points in the bottom panel of Fig.~\ref{fig-bose}, 
which are closest to the LDA curve (i.e., for $L/\xi=10$ and
$V_{\rm max}/\mu_{j=0} < 0.9$, or for $L/\xi=5$ and
$V_{\rm max}/\mu_{j=0} < 0.8$) fall into this regime: the critical
velocity is determined by the energetic instability occurring along
the swallow tails and the critical quasi-momentum is outside the
first Brillouin zone, $P_c > \hbar q_{\rm B}$. Our numerical results
for $P_c$ well agree with those of Ref.~\cite{machholm}.

Finally, from the solution of the GP equation one can calculate the
current(velocity)-phase relation of the superfluid flowing through 
a single potential barrier. Examples are given in the top panel of 
Fig.~\ref{fig-iphi}. The figure shows that, for $L/\xi=10$, the 
results of the GP calculations are close to those in the LDA limit 
not only for the critical velocity (the maximum value of $v$) but 
also for the shape of the curve on the left of the maximum (stable 
branch). On the right of the maximum
(unstable branch) the GP theory still works, providing solitonic
solutions which are instead not accessible to the hydrodynamic
theory without dispersion. We note that, when approaching the
maximum from the right, the amplitude of solitons predicted by the
GP equations vanishes and the two, stable and unstable, branches
merge together \cite{hakim}.

\section{Bogoliubov--de Gennes theory for unitary fermions}

A similar analysis can be done for a superfluid Fermi gas in the
BCS-BEC crossover at zero temperature. A suitable approach consists
in the numerical solution of the BdG equations for the fermionic
quasiparticle amplitudes $u_i$ and $v_i$ \cite{bdg},
\begin{equation}
\left( \begin{array}{cc}
H'(\mathbf r) & \Delta (\mathbf r) \\
\Delta^\ast(\mathbf r) & -H'(\mathbf r) \end{array} \right)
\left( \begin{array}{c} u_i( \mathbf r) \\ v_i(\mathbf r)
\end{array} \right)
=\epsilon_i\left( \begin{array}{c} u_i(\mathbf r) \\
v_i(\mathbf r) \end{array} \right) \; ,
\label{BdGnonuniform}
\end{equation}
with $H'(\mathbf r) =-\hbar^2 \nabla^2/2m +V_{\rm ext}-\mu$. These
equations can be used to calculate the order parameter
$\Delta(\mathbf r) =-g \sum_i u_i(\mathbf r) v_i^*(\mathbf r)$,
the chemical potential $\mu$, and the energy density $e$.

For a superfluid Fermi gas in the presence of a single rectangular
barrier, the BdG equations have already been used by Spuntarelli {\it
et al.} \cite{camerino} to calculate the critical current in the case
of $Lk_{\rm F}=4$. Their results at unitarity are shown in the top
panel of Fig.~\ref{fig-fermi} (dots) where they are compared with the
LDA prediction (\ref{eq:jc}). The figure shows that, for this value of
$Lk_{\rm F}$, the BdG results are indeed close to the LDA curve; in
this regime, also for fermions, the critical velocity is mostly
determined by the local depletion of the density where the potential
is maximum. Concerning the LDA curve, we notice that in the $V_{\rm
max} \to 0$ limit one can easily obtain the analytic expression
\begin{equation}
  v_c/c_s^{(0)} \simeq
1-\sqrt{2}\, \left[V_{\rm max}/(1+\beta)E_{\rm F}\right]^{1/2} .
\end{equation}

In the lower panel of Fig.~\ref{fig-fermi} we instead show our BdG
results for unitary fermions in a periodic potential for various
values of $Lk_{\rm F}$. These results are obtained by using Bloch 
functions and solving the BdG equations with the same self-consistent 
procedure of Ref.~\cite{watanabe}. In particular, the energy density 
$e(n_0,P)$ is calculated as in Eq.~(8) of Ref.~\cite{watanabe} (see 
also \cite{regularization}). From the energy density we then 
obtain the critical quasi-momentum and velocity by means of 
Eq.~(\ref{eq:vcrit-hydro}). 

A striking similarity emerges from Figs.~\ref{fig-bose} and 
\ref{fig-fermi}, revealing that bosons and unitary fermions have
a rather similar behavior. The rapid approach of the BdG results 
to the LDA curve for $Lk_{\rm F}>1$ in the lattice is consistent 
with the case of a single barrier, as well as with the behavior of 
bosons in the corresponding regime, $L/\xi>1$. Our BdG calculations, which 
include an analysis of the quasiparticle energy spectra, show that 
the role of fermionic pair-breaking excitations, not included in
Eq.~(\ref{eq:vcrit-hydro}), is negligible at unitarity, except for
small $V_{\rm max}$ and for very low densities, such that $Lk_{\rm
F}\ll 1$ \cite{note-fermionic}. These fermionic excitations are
instead expected to become important on the BCS side of the
BCS-BEC crossover \cite{combescot,camerino}.

In the bottom panel of Fig.~\ref{fig-iphi} we show the BdG 
results of Ref.~\cite{camerino} for the current(velocity)-phase 
relation of the unitary Fermi superfluid flowing through a single 
potential barrier. As in the case of bosons (top panels) one can 
see that, for $Lk_{\rm F}$ larger than $1$, the BdG results are close
to the LDA limit both for the shape of the stable branch 
and the height of the maximum.

\section{Weakly coupled superfluids}
\label{tb}

By increasing the height of the barriers, such that $V_{\rm
max}\gg \mu$, the critical velocity decreases. In the LDA regime
($L/\xi \gg 1$ or $Lk_{\rm F}\gg 1$) this decrease is fast because
the superfluid density locally follows the shape of the external
potential and rapidly vanishes under a barrier as soon as $V_{\rm
max}$ exceeds $\mu$.  Conversely, when $\xi$ is of order
$L$ or larger, the coupling between the superfluid regions on the
two sides of a barrier remains significant also for larger values
of $V_{\rm max}$, leading to the physics of macroscopic
tunneling and Josephson junctions in weakly coupled superfluids
(for bosons, see for instance \cite{josephson}). 

In this regime, one finds the well-known characteristic sinusoidal 
current-phase relation, i.e., $j(\delta \phi) = j_c \sin(\delta \phi)$, 
where $\delta \phi$ is the phase difference of the superfluid order 
parameter across the barrier \cite{danshita}. In the periodic lattice 
we recover the tight-binding expression for the energy density, which 
is given by $e(n_0,P) = e(n_0,0) + \delta_J [1-\cos(\pi P/P_{\rm edge})]$. 
Here, $P_{\rm edge}$ is the
quasi-momentum at the edge of the Brillouin zone, namely, $P_{\rm
edge}=\hbar q_{\rm B}$ for BECs of bosonic atoms and $P_{\rm
edge}=\hbar q_{\rm B}/2$ for superfluids of fermionic atoms. The
quantity $\delta_J = n_0 P_{\rm edge}^2/\pi^2 m^*$ corresponds to
the half-width of the lowest Bloch band and is related to the 
tunneling energy associated with the Josephson current. The 
effective mass $m^*$ is defined {\it via} the relation $1/m^*
=(1/n_0) \partial_P^2 e(n_0,P)|_{P=0}$ \cite{kramer,book}. 

A consequence of the sinusoidal shape of the energy density in the 
tight-binding limit is that the critical quasi-momentum approaches 
the value $P_c=P_{\rm edge}/2$ and the critical velocity takes the 
following simple dependence on the effective mass:
\begin{equation}
v_c = \frac{1}{\pi}\frac{P_{\rm edge}}{m^*}\, .
\label{eq:tb}
\end{equation}
When $s=V_{\rm max}/E_{\rm R} \gg 1$, the critical velocity decreases
according to  $m/m^* \propto \exp(-2 \sqrt{s})$. The values of $v_c$
obtained from Eq.~(\ref{eq:tb}), with $m^*$ extracted from the GP 
calculation of $e(n_0,P)$, are plotted in the lower panel of 
Fig.~\ref{fig-bose} (thinner black solid lines) 
for $L/\xi=1$ and $1.57$ in the 
region of $V_{\rm max}/\mu_{j=0} \agt 2$. The agreement with the results 
of the complete stability analysis is remarkable. A similar agreement 
is also obtained for unitary fermions, when $e(n_0,P)$ is taken from 
the BdG calculations, as shown in the lower panel of Fig.~\ref{fig-fermi}.

\section{Final remarks}

In all panels of Figs.~\ref{fig-bose} and \ref{fig-fermi} one can
distinguish three limiting cases: i) a regime of hydrodynamic flow
in the local density approximation for large $L/\xi$ (or $Lk_{\rm
F}$), corresponding to the points close to the thick solid lines;
ii) a regime of  macroscopic flow through thin and weak barriers,
where the LDA is not applicable, i.e., for $L/\xi \alt 1$ (or
$Lk_{\rm F} \alt 1$) and $V_{\rm max}/\mu < 1$; iii) a regime of
weakly coupled superfluids separated by thin and strong barriers,
i.e., for $L/\xi \alt 1$ (or $Lk_{\rm F} \alt 1$) but $V_{\rm
max}/\mu \gg 1$.

Concerning bosons, the experiments performed in Ref.~\cite{lens1}
with condensates in deep optical lattices fall into the third
regime and were consistently interpreted in terms of the dynamics
of a chain of Josephson junctions. An investigation of energetic
and dynamical instabilities of bosons in shallow lattices was
instead reported by the same group in Ref.~\cite{lens2}, in a
range of parameters corresponding to the second regime, namely
where the critical velocity for the energetic instability is close to
the sound velocity of the uniform superfluid, as in the standard
Landau criterion. In this experiment, as in the one of
Ref.~\cite{mit_old}, a sharp determination of the Landau critical
velocity was, however, hindered by the inhomogeneity of the trapped
condensate.

\begin{figure}[htbp]
\begin{center}\vspace{0.0cm}
  \rotatebox{0}{\resizebox{8.2cm}{!}
    {\includegraphics{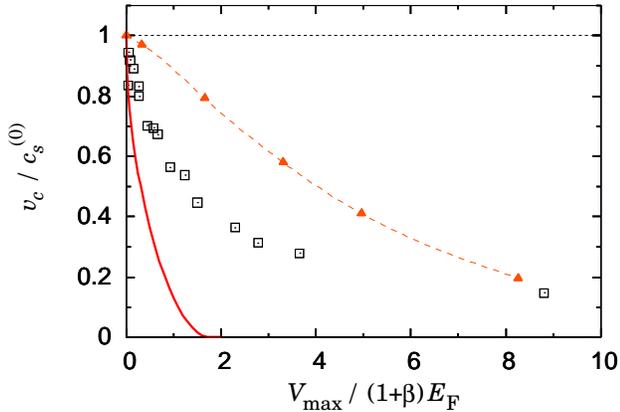}}}
\caption{\label{fig-thexp}(Color online)\quad
Comparison between theoretical results and the experimental data
for the critical velocity of unitary fermions in an optical lattice.
The filled triangles are our BdG results calculated for $E_{\rm F}/E_{\rm R} 
= 0.5$. The solid line is the prediction of hydrodynamics in LDA.
The open squares are the experimental result of Ref.\ \cite{Miller},
normalized by using the density at the $e^{-2}$ waist of the lattice,
which corresponds to $E_{\rm F}/E_{\rm R}=0.56$.
}
\end{center}
\end{figure}

The critical velocity of fermions in an optical lattice is
measured in Ref.~\cite{Miller}. The effects of the inhomogeneity
of the density is reduced by producing the lattice in the central
part of the atomic cloud. However, the lattice itself turns out to
be inhomogeneous. This causes some uncertainties in the
determination of both $k_{\rm F}$ and $V_{\rm max}$, which are
needed for the comparison with a theory for a uniform superfluid
in a uniform lattice. The waist of the two laser beams is smaller,
but not much smaller than the Thomas-Fermi radii of the trapped
gas. A reasonable choice consists of taking the density $n_0$ as
the density measured at the $e^{-2}$ waist of the lattice, as suggested in
Ref.~\cite{Miller}. This yields $E_{\rm F}/E_{\rm R} \simeq 0.56$,
which means $Lk_{\rm F}\simeq 1.1$. In Fig.~\ref{fig-thexp} we
show the experimental data (open squares) for this choice of
$k_{\rm F}$. Our BdG results for $E_{\rm F}/E_{\rm R} =0.5$ are
shown as filled triangles, while the LDA result is represented by the
thick solid line. Here $v_c$ and $V_{\rm max}$ are normalized to
the sound velocity $c_s^{(0)}=\sqrt{(1+\beta)/3}\, v_{\rm F}$ and
the chemical potential $(1+\beta)E_{\rm F}$ of the uniform gas,
respectively \cite{beta}. The behavior of the experimental data 
significantly differs from the theoretical predictions. The disagreement
remains qualitatively the same even for different choices of
$k_{\rm F}$ and $V_{\rm max}$ within the experimental uncertainties 
\cite{note-strength}.  The failure of the LDA is expected, being 
consistent with the fact that here $Lk_{\rm F}$ is of the order of
unity. The disagreement with the BdG calculations is instead puzzling,
especially if one considers that most of the expected inaccuracy of the
BdG theory (e.g., the discrepancy between the mean-field result and the 
Monte-Carlo result of $\beta$) is factorized out by the choice
of the normalization in Fig.~\ref{fig-thexp}. This suggests that 
other effects, including the inhomogeneity of both the density and 
the lattice intensity, as well as the nonstationarity of the 
process, can be important in explaining the experimental observations. 
These effects could be better understood by repeating the same type 
of experiment with bosons, where the GP theory is reliable, full 3D 
GP simulations are feasible and a number of theoretical approaches 
are available to treat the case of time-dependent optical 
lattices (as, for instance, in Bragg spectroscopy experiments), 
from linear response theory to full nonlinear dynamics.    
 
In view of the importance that energetic instabilities 
have in the description of superfluid phenomena, further experimental 
and theoretical investigations on the behavior of ultracold atomic 
superfluids in optical lattices and in the presence of obstacles 
are worth pursuing. An important step is the choice of a suitable 
geometry. Elongated systems with strong transverse confinement 
are good candidates, because their excitation spectrum is simpler. 
A new interesting perspective is also provided by the availability 
of toroidal confining potentials for ultracold gases \cite{nist,franke}, 
in which one can produce a stationary superflow and explore the 
dissipation induced by an external perturbation along the 
torus \cite{piazza}. 

\bigskip

\acknowledgments

We acknowledge M. Modugno and A. Smerzi for fruitful discussions
and for suggestions about the numerical techniques. Calculations
were performed on the HPC facilities WIGLAF at the University of
Trento, BEN at ECT* in Trento, and RIKEN Super Combined Cluster
System. This work, as a part of the European Science Foundation 
EUROCORES Program EuroQUAM-FerMix, is supported by funds
from the CNR and the EC Sixth Framework Programme. It is also
supported by MiUR.

\appendix
\section{Critical velocity in the LDA hydrodynamic theory}
\label{app-lda}

In order to plot the LDA results in Figs.\ \ref{fig-bose} and 
\ref{fig-fermi} we first rewrite Eq.~(\ref{eq:mu}) in terms of the
velocity $v_0\equiv j/n_0$. Its critical value, $v_c$, is determined 
by the condition that the local superfluid velocity $v(z)$ coincides 
with the local sound velocity $c_s(z)$ [Eq.~(\ref{eq:mc2})] at the 
position $z=z_0$, where the potential takes a maximum value 
$V_{\rm ext}(z_0)=V_{\rm max}$. Using this condition, $v(z_0)=
c_s(z_0)=\sqrt{\alpha\gamma n^\gamma(z_0)/m}$, together with the 
continuity equation, $n(z_0)v(z_0)=n_0v_0$, we write Eq.~(\ref{eq:mu}) as
\begin{equation}
\tilde{\mu}_c
=\left(1+\frac{2}{\gamma}\right)\left(\frac{\gamma}{2}\right)^{2/(\gamma+2)}
\tilde{v}_c^{\ 2\gamma/(\gamma+2)}+\tilde{V}_{\rm max},
\label{eq:mu_c}
\end{equation}
where $\tilde{\mu}_c\equiv \mu_{j_c}/\alpha n_0^\gamma$,
$\tilde{v}_c^2\equiv (mv_c^2/2)/\alpha n_0^\gamma$, and 
$\tilde{V}_{\rm max}\equiv V_{\rm max}/\alpha n_0^\gamma$.

In the case of a single potential barrier, one finds $\tilde{\mu}_c=
\tilde{v}_c^2+1$, so that Eq.~(\ref{eq:mu_c}) becomes a third order algebraic 
equation for $\tilde{v}_c^{2/3}$ when $\gamma=1$ (bosons), and a fourth 
order equation for $\tilde{v}_c^{1/2}$ when $\gamma=2/3$ (unitary 
fermions). Once this algebraic equation is solved, one can use 
$\mu_{j=0}=\alpha n_0^\gamma$ [i.e., $\mu_{j=0}=gn_0$ for bosons
and $\mu_{j=0}=(1+\beta)E_{\rm F}$ for unitary fermions] and plot 
$v_c$ as a function of $V_{\rm max}/\mu_{j=0}$.

In the case of the periodic potential the procedure is similar, but 
the chemical potential must be calculated numerically for each value 
of the  average density $n_0$. Equivalently, if $z_1$ is a position 
where $V_{\rm ext}(z_1)=0$, one has $\tilde{\mu}_c=\tilde{n}_1^{-2}
\tilde{v}_c^2+\tilde{n}_1^\gamma$, with $\tilde{n}_1\equiv n(z_1)/n_0$.
We thus obtain an equation for $\tilde{v}_c$ similar to the one of the 
single barrier, but where $\tilde{n}_1$ has to be determined numerically.
Using $\mu_{j=0}$ obtained by a separate calculation, we can 
parametrically plot $v_c$ as a function of $V_{\rm max}/\mu_{j=0}$.

\section{Determination of the critical quasi-momentum in a lattice}
\label{app-pc}

\begin{figure}[htbp]
\begin{center}\vspace{0.0cm}
\rotatebox{0}{
\resizebox{8.2cm}{!}
{\includegraphics{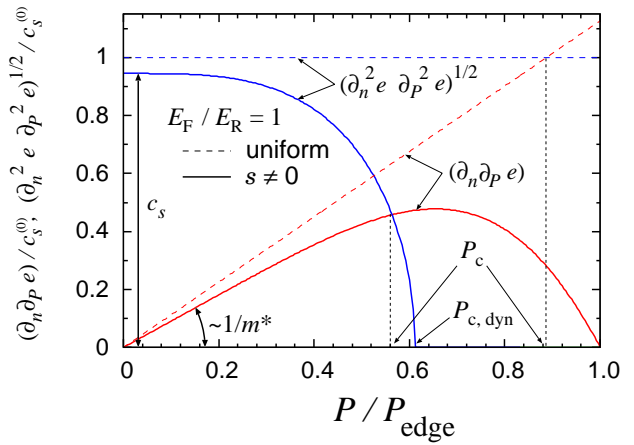}}}
\caption{\label{fig-findpc}(Color online)\quad
Schematic picture for the determination of the critical quasi-momentum 
$P_c$ for the energetic instability using the hydrodynamic relation
(\ref{eq:vcrit-hydro}).  Here we take the unitary Fermi gas
at $E_{\rm F}/E_{\rm R}=1$ as an example. Dashed lines correspond 
to the uniform case ($s=0$) while solid lines to $s=1$. The value
of $P_c$ is given by the abscissa of the crossing point between 
the curve representing $\partial_n\partial_P e$ (red line)
and $(\partial_n^2 e\, \partial_P^2 e)^{1/2}$ (blue line).
The critical quasi-momentum $P_{c,\, {\rm dyn}}$ for the dynamical 
instability of long-wavelength excitations is given by a zero of
$(\partial_n^2 e\, \partial_P^2 e)^{1/2}$, where $\partial_P^2 e=0$.
}
\end{center}
\end{figure}

In this Appendix, we explain how we calculate the critical
quasi-momentum $P_c$ from the relation (\ref{eq:vcrit-hydro}). As
an example, in Fig.~\ref{fig-findpc}, we show the case of the
unitary Fermi gas at $E_{\rm F}/E_{\rm R}=1$, where $P_c$ is in
the first Brillouin zone \cite{note-pc}. Here, we drop the 
subscript ``0'' of the average density $n_0$ for simplicity.

For the uniform case ($s=0$), the energy density can be written as
$e(n,P)=nP^2/2m+e(n,0)$, and thus $\partial_n^2
e(n,P)=\partial_n^2 e(n,0)=\partial_n \mu(n)=1/(n\kappa)$ and
$\partial_P^2 e(n,P)=n/m$. This implies
\begin{align}
\sqrt{\frac{\partial^2 e}{\partial n^2}
\frac{\partial^2 e}{\partial P^2}} =& 
\sqrt{\frac{\kappa^{-1}}{m}}=c_s^{(0)},\\
\frac{\partial^2 e}{\partial n \partial p} =& \frac{P}{m}\ ,
\end{align}
where $\kappa$ is the compressibility and $c_s^{(0)}$ is the sound
velocity of the uniform system: $c_s^{(0)}=\sqrt{(1+\beta)/3}\,
v_{\rm F}$ ($\simeq 0.44 v_{\rm F}$ from the mean-field theory, 
and $\simeq 0.37 
v_{\rm F}$ from Monte Carlo calculations) for unitary fermions and 
$c_s^{(0)}=\sqrt{gn/m}$ for bosons. These results are shown as blue
and red dashed lines in Fig.~\ref{fig-findpc}. Their crossing
point, $P/m=c_s^{(0)}$, gives $P_c$. This procedure is consistent 
with the Landau criterion in a uniform gas.

For the superfluid in a lattice ($s \ne 0$), since $\partial_P
e=0$ at the edge of the Brillouin zone, $\partial_n \partial_P e$
is zero there and has a maximum in the first Brillouin zone (see
the red solid line in Fig.~\ref{fig-findpc}). For the  same reason, 
there exists a new point $P_{c,\, {\rm dyn}}$ at which 
$\partial_P^2 e=0$ and the curve for $\sqrt{\partial_n^2 e\, 
\partial_P^2 e}$ is bent downward (see the blue solid line in 
Fig.~\ref{fig-findpc}).  This $P_{c,\, {\rm dyn}}$, which does not
exist for the uniform case, is the critical quasi-momentum 
for the dynamical instability of long-wavelength excitations; 
the excitation energy $\hbar\omega$ becomes complex for 
$P>P_{c,\, {\rm dyn}}$. The critical quasi-momentum for
the energetic instability $P_c$ is still given by the crossing 
point of $\partial_n\partial_P e$ and $\sqrt{\partial_n^2 e\, 
\partial_P^2 e}$ (red and blue lines) and thus 
$P_c\le P_{c,\, {\rm dyn}}$.

Note that the value of $\sqrt{\partial_n^2 e\, \partial_P^2 e}$ (blue 
curve) at $P=0$ coincides with the sound velocity $c_s$ in the lattice, 
which is decreasing function of $s$ \cite{notectransverse}.  On the 
other hand the slope of $\partial_n\partial_P e$ (red curve) is of 
order $1/m^*$, where $m^*$ is the effective mass.  Since 
$c_s \propto 1/\sqrt{m^*}$ the slope of $\partial_n\partial_P e$ 
approaches zero faster than $c_s$ with increasing $s$. For large 
$s$, the red curve takes a sinusoidal shape and the crossing between 
the two lines occurs closer and closer to $P=P_{\rm edge}/2$. In 
the same limit one finds $P_c=P_{c,\, {\rm dyn}}$. 

It is also worth noticing that, for both bosons
and fermions in the lattice, there exists a value of average
density $n$ such that the critical momentum in the uniform case,
$mc_s^{(0)}$, is equal to the critical quasi-momentum in the 
tight-binding limit, $P_{\rm edge}/2$.  In this case, we have 
numerically checked that $P_c$ is almost independent of the 
lattice strength $s$ in the whole range $s\ge 0$. 
With fermions, this condition occurs when $E_{\rm F}/E_{\rm R}=(3/16)
(1+\beta)^{-1}$; for bosons, it occurs when $gn_0/E_{\rm R}=1/\sqrt{2}$. 
For smaller (larger) density, the critical quasi-momentum $P_c$ 
is a monotonic function of $s$ approaching the asymptotic value 
$P_{\rm edge}/2$ from below (above).

\begin{figure}[htbp]
\begin{center}\vspace{0.0cm}
\rotatebox{0}{
\resizebox{8.2cm}{!}
{\includegraphics{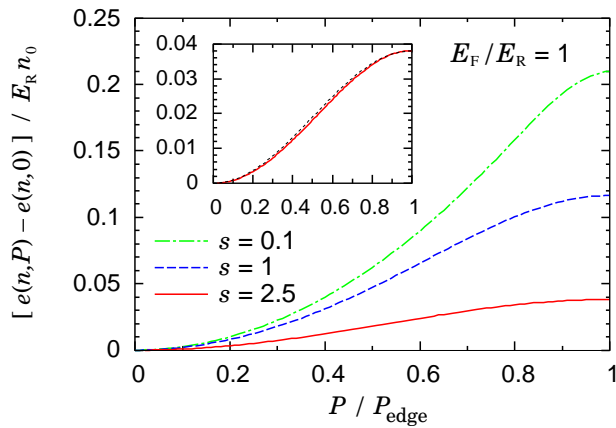}}}
\caption{\label{fig-ep}(Color online)\quad
Energy density for fermions at unitarity as a function of the 
quasi-momentum $P$ for various values of the lattice strength $s$.
Here we take $E_{\rm F}/E_{\rm R}=1$ $(k_{\rm F}L=1.57)$ as an example.
The inset shows the magnification of the result for $s=2.5$;
the dotted line is a sinusoidal curve with the same amplitude.
}
\end{center}
\end{figure}

\section{Band structure of the superfluid unitary Fermi gases in 
periodic potentials}
\label{app-band}

For superfluid Fermi gases in an optical lattice,
solving the BdG equations with a large energy cutoff $E_{\rm C}$
is rather time-consuming.  In the present work,
in order to obtain the $P$-dependence of $e$ and $\mu$ for various 
parameters, we calculate them for a limited number of 
points (typically seven or more) in the first Brillouin zone for 
each $E_{\rm F}/E_{\rm R}$ and we use the following fitting function:
\begin{align}
  f(n_0,P)=&f(n_0,0)\nonumber\\
&+ \delta(n_0)
\left\{\frac{1}{2}\left(1-\cos{\left[\pi\left(\frac{P}{P_{\rm
edge}}\right)^\eta\ \right]}\right) \right\}^{1/\eta},
\label{eq:fit}
\end{align}
with the band width,
\begin{equation}
\delta(n_0)\equiv f(n_0, P_{\rm edge})-f(n_0,0).
\end{equation}
The shape parameter $\eta$ can take all values 
form $1$ to $\infty$: the case $\eta=1$ corresponds to
the sinusoidal form for strong lattices, while 
$\eta=\infty$ reproduces the quadratic energy 
dispersion of the translationally invariant system
without lattice ($s=0$). Note that $\partial_P f(n_0,P)=0$ 
at $P=0$ and $P_{\rm edge}$ like $e$ and $\mu$ (here, we do 
not consider situations where the band structure exhibits 
``swallow tails''). We find that the above fitting function 
(\ref{eq:fit}) very well reproduces the numerical BdG results 
of $e$ and $\mu$ except for the cases of very small $s$. Even 
for $s=0.1$, which is the smallest non-zero value studied in
the present work, the relative error is typically less 
than $0.1\%$. 

In Fig.~\ref{fig-ep}, we show the energy density $e(n,P)$ of
the unitary Fermi superfluid in a periodic potential 
calculated by the BdG equations for $E_{\rm F}/E_{\rm R}=1$
and for three values of the lattice strength, $s=0.1$, 
$1$, and $2.5$. The corresponding values of the shape 
parameter $\eta$ are $\eta=5.05$, $1.68$, and $1.13$. 
Note that the $P$-dependence of $e$ is rather close to a 
quadratic form for the weakest lattice ($s=0.1$) and is almost 
sinusoidal for the strongest lattice ($s=2.5$), as one 
can see in the inset of Fig.~\ref{fig-ep}.


\begin{thebibliography}{99}
\bibitem{kink} We observe that, in the LDA hydrodynamic theory,
$n(z)$ and $v(z)$ exhibit a kink at the point where $v(z)=c_s(z)$,
with a finite jump in the first derivative, and one cannot
construct a stationary solution for $v(z)>c_s(z)$.
\bibitem{smoothing} Whenever the critical condition for Landau instability 
is reached at a position $z_0$ far from the barrier edge, at a distance 
much larger than the healing length $\xi$, the results for the critical 
velocity are insensitive to the details of the shape of the barrier and 
are expected to be the same for both sharp and smooth barrier. In the 
LDA limit, this condition is satisfied and, indeed, our results for the 
rectangular barrier are consistent with the results discussed in 
Ref.\ \cite{hakim} 
for a generic slowly varying potential. Of course, the LDA would not be 
reliable in predicting, for example, the behavior of the density distribution 
near the edge of the barrier, within a distance of order $\xi$.
\bibitem{solvingGPE} Given $j$ and $n_0$, Eq.~(\ref{eq:gp}) admits two
symmetric bounded stationary solutions which merge and disappear at
the critical point $j=j_c$. These solutions can be expressed in terms
of Jacobi elliptic functions \cite{mamaladze} in which $j$ and $n_0$
enter as parameters. Upon matching the density derivative at the edge
of the barrier, one is left with only one free parameter, to be in
turn fixed by the matching of the density itself. The latter gives a
non-algebraic equation, which we solve by means of a standard
numerical root finding method.
\bibitem{note-fermionic} The creation of fermionic pair-breaking
excitations under a barrier implies a transfer of momentum
to the barrier of the order of $\sim 2\hbar k_F$. The probability
of this process is expected to be proportional to $\exp(-Lk_F)$ and
hence to be negligible for large $L$.
\bibitem{note-strength} In Fig.~\ref{fig-thexp}, 
we use the same lattice height used in 
Ref.~\cite{Miller}, which is characterized by the peak value over the sample.
If we use the lattice height at the $e^{-2}$ waist instead, 
$V_{\rm max}$ of both the experimental data (open squares) 
and our theoretical prediction (filled squares) will be reduced 
by the same factor and thus the discrepancy still remains.
\bibitem{note-pc} In general, there are also cases where $P_c$ is outside 
the first Brillouin zone, as we have seen for bosons when the energy 
exhibits swallow tails. In Appendix B, we do not consider this case for
simplicity, but the procedure for calculating $P_c$ is the same.
\bibitem{notectransverse} Note that in this work we always
consider a flow along the direction of the 1D lattice. It is worth 
noticing here that the critical velocity of the superfluid flowing instead 
along the transverse direction can be larger than the sound velocity for 
the uniform system $c_s^{(0)}$. In fact, since the system is translationally 
invariant in the transverse direction, one has $m^*=m$, 
$\partial_n\partial_{P_\perp} e=P_{\perp}/m$, and $\kappa$ does not depend 
on the transverse momentum $P_{\perp}$. On the other hand, due to the 
periodic potential, $\kappa^{-1}$ can be larger than that of the uniform 
system with the same average density \cite{watanabe,kramer}.
Under these conditions, both the sound velocity in the lattice and
the critical velocity, $v_c = \sqrt{\partial_n^2 e\, \partial_P^2 e}
=\sqrt{\kappa^{-1}/m}$, are larger than  $c_s^{(0)}$.
\end{thebibliography}
\end{document}